\documentclass[conference]{IEEEtran}
\IEEEoverridecommandlockouts

\usepackage{cite}
\usepackage{amsmath,amssymb,amsfonts}
\usepackage{booktabs}
\usepackage{multirow}
\usepackage{url}
\usepackage{caption}
\usepackage{microtype}
\usepackage[T1]{fontenc}
\usepackage[utf8]{inputenc}
\usepackage{textcomp}

\usepackage{flushend}

\begin{document}

\title{Zero-Shot Voice Cloning\\
for Synthetic Data Augmentation in Children's ASR}

\author{
  \IEEEauthorblockN{Mohammad Mohsen Goodarzi,
                    Stefan Wabnik,
                    Matthias Stennes,\\
                    Stephanus Volke,
                    Alexander Schwarz}
  \IEEEauthorblockA{Fraunhofer Institute for Digital Media Technology IDMT,\\ Oldenburg
Branch for Hearing, Speech and Audio Technology HSA, Oldenburg,
Germany\\
    \texttt{\{mohammadmohsen.goodarzi, stefan.wabnik, matthias.stennes,}\\
    \texttt{stephanus.volke, alexander.schwarz\}@idmt.fraunhofer.de}}

}

\maketitle

\begin{abstract}
Children's automatic speech recognition (ASR) lags behind adult ASR due to the
acoustic mismatch between adult-trained acoustic models and children's speech, as
well as the scarcity of labelled children's speech corpora.
We investigate whether zero-shot text-to-speech (TTS) synthesis with voice
cloning, requiring only a short reference clip per speaker, can serve as a
practical substitute for real children's speech when fine-tuning large
pre-trained ASR models.
Using the OmniVoice multilingual TTS system we synthesise up to 16\,000
child-like utterances from two English corpora (ENNI, clinical narrative samples,
ages 7--12, and MyST, conversational science-tutor speech, grades 3--5) and
fine-tune Whisper-small under different experimental conditions covering real data,
synthetic data, cross-corpus data, and all-data combination.
A systematic sensitivity analysis over speaker count, text-pool size, and utterances per
speaker reveals that synthetic data alone reduces word error rate (WER) from a
19.8\,\% zero-shot baseline to 16.4\,\% on ENNI, recovering approximately 30\,\%
of the gap to real-data fine-tuning (8.8\,\%).  Combining synthetic with real data is
essentially indistinguishable from real-data-only fine-tuning (9.2\,\% vs.\ 8.8\,\%).
Speaker count plateaus at ${\approx}40$ speakers for cross-domain transfer, while
multi-speaker acoustic diversity per transcript, rather than raw text
coverage, emerges as the primary determinant of cross-domain generalisation;
out-of-vocabulary (OOV) rate proves to be a poor proxy for WER given Whisper's
pre-trained language modelling capacity.
\end{abstract}


\section{Introduction}

Automatic speech recognition systems have reached near-human performance on
adult read speech benchmarks~\cite{radford2023whisper}, yet children's ASR
remains a substantially harder problem.  Children differ from adults in their
fundamental frequency (higher F0, broader range), formant frequencies, speaking
rate, and co-articulation patterns; they also exhibit more disfluencies and
reduced phonetic precision, all of which cause ASR systems to produce higher
word error rates on children's speech relative to
text~\cite{gerosa2009review,potamianos2003robust}.
At the same time, labelled children's speech corpora are orders of magnitude
smaller than adult corpora: collecting clinical-grade narrative samples requires
specialist annotation, and freely distributable children's data is scarce due to
privacy regulations.

Large self-supervised and weakly-supervised models such as
wav2vec~2.0~\cite{baevski2020wav2vec} and Whisper~\cite{radford2023whisper} have
dramatically reduced the data requirements for adult ASR by providing powerful
pre-trained representations.  When fine-tuned on even small children's datasets
they show significant improvement~\cite{jain2023wav2vec2children,attia2024kidwhisper}.
However, the adaptation gains are proportional to the amount of in-domain
fine-tuning data, and obtaining more children's speech remains costly.

A natural response is data augmentation via TTS synthesis.  Prior work explored
trainable children's TTS~\cite{zhang2024childaug}, voice conversion from adult
speech~\cite{shahnawazuddin2020transfer}, and acoustic feature
perturbation~\cite{potamianos2003robust}.  The recent generation of zero-shot
voice cloning models, systems that require only a few seconds of reference audio
at \emph{inference} time, with no speaker-specific training or
enrolment~\cite{casanova2022yourtts,wang2023valle,zhu2026omnivoice}, opens
a qualitatively new possibility: extract reference clips directly from an
existing corpus, clone each speaker's voice, and synthesise as much additional
text in that voice as desired, without training a separate TTS system.
We use the term ``zero-shot'' in the sense standard in the TTS literature:
the model is not trained or adapted on the target speaker; the reference clip
serves only as a conditioning prompt at synthesis time.

We make the following contributions:
\begin{enumerate}
\item We present an end-to-end zero-shot voice cloning pipeline for children's
  ASR data augmentation that requires only a single short reference clip per
  child speaker, with no TTS model retraining, and demonstrate its
  effectiveness using OmniVoice~\cite{zhu2026omnivoice} across two structurally
  distinct English children's corpora.
\item Through a comprehensive comparison of different training conditions spanning
  real-only, synthetic-only, cross-corpus, and combined configurations, we show
  that synthetic data alone recovers approximately 30\,\% of the gap to
  real-data fine-tuning, while combining synthetic and real data yields
  performance statistically indistinguishable from real-only fine-tuning.
\item We conduct a systematic sensitivity analysis over synthesis budget
  factors (speaker count 5--80, text-pool coverage 10--100\,\%, and
  utterances per speaker 25--200), revealing that multi-speaker acoustic
  coverage per transcript, rather than raw text diversity or speaker count alone,
  is the primary driver of cross-domain generalisation; that speaker count
  plateaus at approximately 40 for out-of-domain transfer; and that 50--100
  utterances per speaker suffice for near-saturation in-domain performance.
\end{enumerate}

\section{Related Work}

\subsection{Children's ASR}

Early children's ASR relied on acoustic model adaptation using
maximum-likelihood or feature-space transforms~\cite{potamianos2003robust}.
The CSLU Kids corpus~\cite{shobaki2000ogikids} and the My Science Tutor (MyST)
corpus~\cite{pradhan2021myst} provided the first large-scale benchmarks.
More recent work has focused on fine-tuning self-supervised models on children's
speech: Jain et al.~\cite{jain2023wav2vec2children} fine-tuned wav2vec~2.0
models on MyST and demonstrated large WER reductions; Attia et
al.~\cite{attia2024kidwhisper} (Kid-Whisper) showed that fine-tuning Whisper on
MyST closes much of the adult--child gap.  Singh et al.~\cite{singh2026causal}
provided a causal analysis of the factors driving children's ASR degradation,
quantifying the contributions of physiological, cognitive, and extrinsic factors.

\subsection{TTS-Based Data Augmentation for ASR}

Rosenberg et al.~\cite{rosenberg2019ttsasr} were among the first to demonstrate
that TTS-generated speech can improve ASR training, particularly for low-resource
conditions.  V\'{a}squez-Correa et al.~\cite{vasquez2023whispertts} showed that
Whisper models can be domain-adapted using only synthetic speech.  For
children's ASR specifically, Zhang et al.~\cite{zhang2024childaug} augmented
child-like speech via voice conversion and showed WER reductions for
Whisper-based systems.  Rolland and Abad~\cite{rolland2025personalised}
demonstrated per-child personalisation via TTS voice cloning, achieving
significant WER reduction for 10 target children.  However, their approach
requires enrolment speech from each test child and produces one adapted model
per child.  In contrast, we investigate whether voice cloning of training-set
children, requiring \emph{no test-child enrolment}, can improve
speaker-independent ASR for arbitrary unseen children, and conduct the first
systematic analysis of the synthesis budget factors (speaker count, text
diversity, and utterances per speaker) that drive this improvement.

\subsection{Zero-Shot TTS and Voice Cloning}

Modern zero-shot TTS systems can clone a speaker's voice from a 3--10\,s
reference clip, enabling on-demand speaker expansion of any text corpus.
VITS~\cite{kim2021vits} introduced a fully end-to-end variational approach
that underpins many subsequent systems.  YourTTS~\cite{casanova2022yourtts}
extended VITS to zero-shot multi-speaker and zero-shot voice conversion settings.
VALL-E~\cite{wang2023valle} proposed a neural codec language model that preserves
speaker emotion and environment from a short prompt.
OmniVoice~\cite{zhu2026omnivoice}, the system used in this work, is a
diffusion language-model-based architecture trained on 581\,k hours of
multilingual open data that supports over 600 languages via a
non-autoregressive full-codebook token prediction scheme, achieving
state-of-the-art speaker similarity and intelligibility.

\section{Methodology}

\subsection{Corpora}

\textbf{ENNI} (Edmonton Narrative Norms Instrument)~\cite{schneider2006enni} is
a clinical corpus of English narrative uttered by children aged 7--12 using
picture sequences.  We use only the control (TD) subset.  After standard
train/development/test splits and filtering, we retain 10\,328 training utterances
(9.5\,h, 116 speakers), 1\,257 development utterances (1.2\,h, 14 speakers), and
1\,278 test utterances (1.1\,h, 15 speakers).

\textbf{MyST} (My Science Tutor)~\cite{pradhan2021myst} is a large corpus of
conversational science tutoring sessions with elementary school children (grades
3--5).  We use 75\,324 training utterances (167.8\,h, 567 speakers), 11\,963
development utterances (25.3\,h), and 12\,915 test utterances (27.9\,h, 91 speakers).

\subsection{Synthesis Pipeline}

\textbf{Persona extraction.}
For each corpus we select up to 80 training speakers as \emph{personas} for
voice cloning.  For each persona we identify a single reference clip: the
utterance closest to 4\,s duration that is at least 2\,s long.  The clip and
its transcription form the persona's voice seed.

\textbf{Text normalisation.}
Source transcriptions are normalised via a shared pipeline: CHAT annotation
markup, disfluency markers, noise tokens (\texttt{xxx}/\texttt{yyy}/\texttt{www}),
and punctuation are removed; text is lowercased and whitespace-normalised.
Filler words (\emph{um}, \emph{uh}, \emph{hmm}) are removed so the TTS model
does not produce scripted hesitations.

\textbf{Source filtering.}
Records are excluded from the synthesis pool if: (a) the source utterance exceeds
20\,s, (b) the normalised text contains more than 30 words, or (c) the text is a
duplicate.

\textbf{Synthesis.}
The filtered, deduplicated text pool is passed to
OmniVoice~\cite{zhu2026omnivoice} (\texttt{k2-fsa/OmniVoice}, April 2026
release), which conditions on each persona's reference clip.  We synthesise up
to $N$ utterances per persona at 16\,kHz, parallelised across four NVIDIA GPUs.

\textbf{Post-synthesis filtering.}
Each clip is validated with a heuristic audio--text plausibility check:
utterances are rejected if the inferred speaking rate falls outside 20--600\,WPM,
or if the same word repeats for ten or more consecutive tokens.  Surviving clips
are indexed in JSONL manifests split 80\,\%/10\,\%/10\,\% train/development/test by
persona.  Default synthesis produces 16\,000 training utterances
(80 personas $\times$ 200 utterances each).

Table~\ref{tab:datasets} summarises the training datasets used across the main
experiments.

\begin{table}[t]
  \caption{Training dataset statistics. Synthetic sets use 80 child voice
    personas (200\,utt/persona). Speaker counts for synthetic sets refer to
    the number of voice personas cloned from the source corpus.}
  \label{tab:datasets}
  \centering
  \scriptsize
  \begin{tabular}{@{}llrrr@{}}
    \toprule
    Dataset & Type & Utt. & Dur.\ (h) & Spk. \\
    \midrule
    ENNI-Real            & Real (ENNI)                & 10\,328 & 9.5  & 116 \\
    MyST-Real            & Real (MyST)                & 75\,324 & 167.8 & 567 \\
    ENNI-Syn             & Synth.\ (ENNI spk/txt)     & 16\,000 & 16.6 & 80  \\
    MyST-Syn             & Synth.\ (MyST spk/txt)     & 15\,994 & 25.8 & 80  \\
    MyST-Syn[ENNI-Text]  & Synth.\ (MyST spk, ENNI txt)& 15\,998 & 17.8 & 80  \\
    ENNI-Syn[MyST-Text]  & Synth.\ (ENNI spk, MyST txt)& 16\,000 & 24.3 & 80  \\
    \bottomrule
  \end{tabular}
\end{table}

\subsection{ASR Model and Training}

We fine-tune \textbf{Whisper-small}~\cite{radford2023whisper} (244\,M
parameters), with language set to \texttt{english} and
\texttt{task=transcribe}.  Training uses the HuggingFace
\texttt{Seq2SeqTrainer} with: batch size 32, gradient accumulation 2 (effective
batch 64), learning rate $10^{-5}$, cosine schedule with 500 warm-up steps,
maximum 4\,000 steps, FP16.  Features are extracted at 16\,kHz and
padded/truncated to 30\,s.  We use 4-GPU data-parallel training via HuggingFace
Accelerate.

\subsection{Evaluation}

WER is computed over normalised text (lowercase, punctuation stripped) on four
test sets: \textbf{enni\_test} (real children, in-domain for ENNI-trained
systems), \textbf{myst\_test} (real children, cross-domain),
\textbf{synthetic\_enni\_test} (4\,000 synthesised utterances from ENNI
personas), and \textbf{synthetic\_myst\_test} (4\,000 synthesised utterances
from MyST personas).  Confidence intervals (CIs) are estimated via
non-parametric bootstrap resampling (1\,000 samples, 95\,\% CI).  We additionally
report the out-of-vocabulary (\textbf{OOV}) rate: the fraction of test-set word
tokens absent from the training vocabulary.

\section{Results and Analysis}

\subsection{Real vs.\ Synthetic Data}

Table~\ref{tab:main} summarises the main experimental results.  The zero-shot
Whisper-small baseline achieves 19.8\,\% WER on \texttt{enni\_test} and
22.2\,\% on \texttt{myst\_test}.  Fine-tuning on real ENNI data (E1) reduces
\texttt{enni\_test} WER to 8.8\,\% but shows no generalisation (MyST WER
$=21.7$\,\%, nearly unchanged from baseline), as the model specialises to ENNI
vocabulary.  Synthetic ENNI data (E2) yields 16.4\,\% on \texttt{enni\_test}:
it bridges approximately 30\,\% of the real-data gap while maintaining similar
\texttt{myst\_test} WER (21.0\,\%).

Combining real and synthetic ENNI (E7) achieves 9.2\,\% on \texttt{enni\_test},
which is within the 95\,\% CI of real-only fine-tuning ($8.8 \pm 0.8$\,\%), and
20.6\,\% on \texttt{myst\_test}, indicating that synthetic data can help in
generalisation without hurting in-domain performance.  The all-data condition
(E9) achieves 8.5\,\% on \texttt{enni\_test} and 14.8\,\% on
\texttt{myst\_test}, the best overall result.

The cross-dataset experiments (E10--E11) are instructive: using MyST speakers
with ENNI text (E10) achieves 17.7\,\% on \texttt{enni\_test}, while using ENNI
speakers with MyST text (E11) degrades to 24.5\,\%.  This asymmetry indicates
that, in our setting, \textbf{matching the text domain of the target corpus
contributes more than speaker voice identity} to in-domain WER.

\begin{table*}[t]
  \caption{Main experiment results. WER\,(\%) shown as mean $\pm$ half 95\,\% CI.
    OOV rate\,(\%) in parentheses.
    Bold = best real-children WER per column.}
  \label{tab:main}
  \centering
  \small
  \begin{tabular}{@{}clcccc@{}}
    \toprule
    Exp & Training data & ENNI-Real WER & MyST-Real WER & ENNI-Syn WER & MyST-Syn WER \\
    \midrule
    E0  & --- (zero-shot)                                & $19.8 \pm 1.1$        & $22.2 \pm 1.4$        & $7.7 \pm 0.4$  & $8.2 \pm 2.1$  \\
    E1  & ENNI-Real                                      & $\mathbf{8.8 \pm 0.8}$ (1.1) & $21.7 \pm 1.3$ (21.2) & $4.4 \pm 0.3$  & $7.3 \pm 1.6$  \\
    E2  & ENNI-Syn                                       & $16.4 \pm 1.1$ (1.2)  & $21.0 \pm 0.7$ (21.7) & $2.1 \pm 0.2$  & $4.8 \pm 0.3$  \\
    E3  & MyST-Real                                      & $17.7 \pm 1.1$ (4.1)  & $\mathbf{14.6 \pm 0.9}$ (0.4) & $7.3 \pm 0.3$ & $4.4 \pm 1.1$ \\
    E4  & MyST-Syn                                       & $26.5 \pm 1.2$ (8.7)  & $19.0 \pm 0.9$ (1.2)  & $5.7 \pm 0.3$  & $2.6 \pm 0.2$  \\
    E5  & ENNI-Real + MyST-Real                          & $13.1 \pm 6.7$ (0.4)  & $14.6 \pm 0.8$ (0.4)  & $4.3 \pm 0.2$  & $3.8 \pm 0.2$  \\
    E6  & ENNI-Syn + MyST-Syn                            & $17.7 \pm 1.1$ (0.7)  & $18.7 \pm 0.7$ (1.1)  & $2.2 \pm 0.2$  & $2.6 \pm 0.2$  \\
    E7  & ENNI-Real + ENNI-Syn                           & $9.2 \pm 0.8$ (1.1)   & $20.6 \pm 0.7$ (21.2) & $1.9 \pm 0.2$  & $5.6 \pm 0.3$  \\
    E8  & MyST-Real + MyST-Syn                           & $21.3 \pm 6.8$ (4.1)  & $16.1 \pm 1.3$ (0.4)  & $6.6 \pm 1.2$  & $2.9 \pm 0.2$  \\
    E9  & ENNI-Real + MyST-Real + ENNI-Syn + MyST-Syn   & $\mathbf{8.5 \pm 0.8}$ (0.4) & $\mathbf{14.8 \pm 0.9}$ (0.4) & $2.4 \pm 0.2$ & $2.6 \pm 0.2$ \\
    E10 & MyST-Syn[ENNI-Text]                            & $17.7 \pm 1.2$ (1.3)  & $21.7 \pm 1.0$ (21.5) & $2.4 \pm 0.2$  & $4.5 \pm 0.3$  \\
    E11 & ENNI-Syn[MyST-Text]                            & $24.5 \pm 1.3$ (6.9)  & $20.6 \pm 1.4$ (1.1)  & $5.6 \pm 0.3$  & $3.2 \pm 0.2$  \\
    \bottomrule
  \end{tabular}
\end{table*}

\subsection{Speaker Count Sensitivity}

Speaker diversity shows diminishing returns for in-domain performance: all
speaker-count sensitivity (SA) conditions (5--80 speakers) land in
17.0--17.9\,\% on \texttt{enni\_test} with overlapping CIs, indicating that text
and acoustic coverage from 16\,k utterances dominates over speaker count for the
matched test corpus.

For \textbf{cross-domain} generalisation (\texttt{myst\_test}), speaker count
matters more: SA\_5 achieves $22.6 \pm 0.9$\,\%, while SA\_40 achieves
$20.8 \pm 0.8$\,\%, a 1.8\,pp improvement whose CIs do not overlap.  SA\_80
($21.3 \pm 0.7$\,\%) is within CI of SA\_40, confirming that going beyond 40
speakers offers no additional benefit when total utterances are fixed.

\begin{table}[t]
  \caption{Speaker count sensitivity. All conditions use ENNI text pool (100\,\%)
    with utterances distributed evenly across personas. OOV rate\,(\%) in parentheses.}
  \label{tab:speaker}
  \centering
  \scriptsize
  \begin{tabular}{@{}ccrcc@{}}
    \toprule
    Exp & Spk. & Total utt. & ENNI-Real WER & MyST-Real WER \\
    \midrule
    SA\_5  & 5  & 16\,000 & $17.4 \pm 1.1$ (1.1) & $22.6 \pm 0.9$ (21.6) \\
    SA\_10 & 10 & 16\,000 & $17.9 \pm 1.1$ (1.2) & $22.8 \pm 1.0$ (21.6) \\
    SA\_20 & 20 & 16\,000 & $17.7 \pm 1.1$ (1.3) & $21.4 \pm 0.7$ (21.7) \\
    SA\_40 & 40 & 16\,000 & $17.0 \pm 1.0$ (1.2) & $20.8 \pm 0.8$ (21.5) \\
    SA\_80 & 80 & 16\,000 & $17.0 \pm 1.1$ (1.2) & $21.3 \pm 0.7$ (21.7) \\
    \bottomrule
  \end{tabular}
\end{table}

\subsection{Text-Pool Sensitivity}

We vary the fraction of ENNI training transcriptions available as synthesis text
prompts while fixing 80 speakers and 100 utterances per speaker (8\,000
utterances total).  With only 10\,\% of the text pool (967 unique texts), TA\_10
achieves the lowest WER on both \texttt{enni\_test} (16.6\,\%) and
\texttt{myst\_test} (20.0\,\%), despite having the highest OOV rates (4.3\,\%
and 29.0\,\% respectively).  The TA\_10 vs.\ TA\_100 gap on \texttt{myst\_test}
(20.0 vs.\ 20.9) is at the boundary of statistical significance (CIs barely
overlap).  The likely cause is that each of the 967 texts was synthesised by
${\approx}8.3$ different speakers on average (vs.\ ${\approx}1.5\times$ repetition
in TA\_100), providing richer multi-speaker acoustic variability per
transcript, a form of implicit multi-speaker data augmentation.

Increasing text budget beyond 10\,\% produces no consistent WER improvement.
OOV rate is a poor proxy for WER in this setting: TA\_10 has the highest OOV
rate yet the best WER, because Whisper's decoder has broad pre-trained language
modelling capacity.

\begin{table}[t]
  \caption{Text-pool sensitivity. All conditions use 80 ENNI personas $\times$
    100\,utt each. OOV rate\,(\%) in parentheses.}
  \label{tab:textpool}
  \centering
  \scriptsize
  \begin{tabular}{@{}ccrcc@{}}
    \toprule
    Exp & Text pool & Total utt. & ENNI-Real WER & MyST-Real WER \\
    \midrule
    TA\_10  & 10\,\%  & 8\,000 & $16.6 \pm 1.0$ (4.3)  & $20.0 \pm 0.7$ (29.0) \\
    TA\_25  & 25\,\%  & 8\,000 & $18.0 \pm 1.1$ (2.5)  & $20.7 \pm 0.8$ (24.6) \\
    TA\_50  & 50\,\%  & 8\,000 & $17.2 \pm 1.1$ (1.7)  & $20.9 \pm 1.0$ (23.3) \\
    TA\_100 & 100\,\% & 8\,000 & $17.6 \pm 1.1$ (1.5)  & $20.9 \pm 0.6$ (22.7) \\
    \bottomrule
  \end{tabular}
\end{table}

\subsection{Utterances-per-Speaker Sensitivity}

With 80 speakers and the full ENNI text pool, we vary the number of synthesised
utterances per persona: 25, 50, 100, 200.  ENNI-Real WER decreases rapidly from
18.4\,\% (25\,utt/speaker) to 17.0\,\% (50\,utt/speaker), then plateaus at
17.0--17.5\,\% for 100--200\,utt/speaker.  \textbf{Synthesis of approximately
50--100 utterances per speaker saturates in-domain WER improvement}, making it a
cost-effective operating point.  Adding more utterances for the same 80 speakers
slightly hurts out-of-domain generalisation, possibly reflecting increased
over-fitting to speaker-specific TTS artefacts.

\begin{table}[t]
  \caption{Utterances-per-speaker sensitivity. All conditions use 80 ENNI personas
    and the full text pool. OOV rate\,(\%) in parentheses.}
  \label{tab:utt}
  \centering
  \scriptsize
  \begin{tabular}{@{}ccrcc@{}}
    \toprule
    Exp & Utt/spk & Total utt. & ENNI-Real WER & MyST-Real WER \\
    \midrule
    UA\_25  & 25  & 2\,000  & $18.4 \pm 1.1$ (2.9) & $22.5 \pm 1.2$ (25.4) \\
    UA\_50  & 50  & 4\,000  & $17.0 \pm 1.1$ (2.0) & $21.7 \pm 0.8$ (24.1) \\
    UA\_100 & 100 & 8\,000  & $17.5 \pm 1.1$ (1.5) & $21.4 \pm 0.7$ (22.3) \\
    UA\_200 & 200 & 16\,000 & $17.0 \pm 1.2$ (1.2) & $22.1 \pm 0.8$ (21.7) \\
    \bottomrule
  \end{tabular}
\end{table}

\subsection{Speakers-per-Text Sensitivity}

The speakers-per-text (SPT) analysis decouples speaker diversity from text
diversity: with 8\,000 utterances and 80 speakers, we vary how many speakers
read the same text item.  When each text is read by only 1 speaker (SPT\_1,
83\,\% text pool), ENNI-Real WER is 16.8\,\% with an OOV of 1.6\,\%.  When every
text is read by all 80 speakers (SPT\_80, 1\,\% text pool), WER rises modestly to
17.5\,\% but OOV explodes to 18.8\,\%.

Crucially, \texttt{myst\_test} WER does \textbf{not} increase with higher OOV:
SPT\_8 (10\,\% text pool, OOV 29.0\,\%) achieves the best cross-domain WER of
19.8\,\%, outperforming SPT\_1 (83\,\% text pool, OOV 22.7\,\%) at 20.9\,\%.
The OOV--WER relationship on \texttt{myst\_test} is \textbf{not monotonic}; it
is weakly U-shaped, with a minimum near SPT\_8.  The results support the
conclusion that \textbf{multi-speaker acoustic diversity per transcript is the
primary driver of cross-domain generalisation}: having 8 speakers produce the
same text provides richer acoustic variation for the encoder to learn from.
Beyond SPT\_8, further reduction in text pool (SPT\_16, SPT\_80) yields
diminishing returns.

\begin{table}[t]
  \caption{Speakers-per-text sensitivity. All conditions use 80 ENNI personas,
    8\,000 utterances total. OOV rate\,(\%) in parentheses.}
  \label{tab:spt}
  \centering
  \scriptsize
  \begin{tabular}{@{}ccrccc@{}}
    \toprule
    Exp & Spk/txt & Text pool & ENNI-Real WER & MyST-Real WER \\
    \midrule
    SPT\_1  &  1 & 83\,\% & $16.8 \pm 1.0$ (1.6)  & $20.9 \pm 0.7$ (22.7) \\
    SPT\_2  &  2 & 41\,\% & $17.2 \pm 1.1$ (1.7)  & $21.1 \pm 0.8$ (23.4) \\
    SPT\_4  &  4 & 21\,\% & $17.4 \pm 1.0$ (2.8)  & $20.7 \pm 0.6$ (25.2) \\
    SPT\_8  &  8 & 10\,\% & $16.8 \pm 1.0$ (4.3)  & $\mathbf{19.8 \pm 0.5}$ (29.0) \\
    SPT\_16 & 16 &  5\,\% & $16.8 \pm 1.0$ (6.5)  & $20.2 \pm 0.5$ (32.7) \\
    SPT\_80 & 80 &  1\,\% & $17.5 \pm 1.0$ (18.8) & $20.4 \pm 0.6$ (42.4) \\
    \bottomrule
  \end{tabular}
\end{table}

\subsection{Synthetic vs.\ Real Evaluation Gap}

An important confound is that evaluation on \emph{synthetic} test sets (4\,000
clips from held-out personas) yields much lower WER than real test sets, e.g.\
E2 gives 2.1\,\% on \texttt{synthetic\_enni\_test} but 16.4\,\% on real
\texttt{enni\_test}.  As expected, evaluating a synthetically trained model on
matched synthetic data does not suffer a domain gap; we report these numbers
mainly to quantify the magnitude of that mismatch.  The synthetic-to-real gap
decreases when real data is added: E7 gives 1.9\,\% synthetic vs.\ 9.2\,\% real.
This confirms that \textbf{synthetic evaluation benchmarks substantially
overestimate ASR quality on real children's speech} and should always be paired
with real-children evaluations.

\section{Conclusion}

We have demonstrated that zero-shot voice cloning with OmniVoice provides a
practical path to synthesising children's speech for ASR fine-tuning from as few
as 5 speaker reference clips (one clip per speaker).  Synthetic data alone
recovers ${\approx}30$\,\% of the gap to real-data fine-tuning; combined with
real data it is statistically indistinguishable from real-only fine-tuning on
ENNI.  The sensitivity analysis yields three actionable insights:

\begin{itemize}
\item \textbf{Speaker count:} plateau at ${\approx}40$ speakers for cross-domain
  generalisation; 5 speakers suffice for in-domain saturation.
\item \textbf{Acoustic diversity per transcript:} the primary driver of
  cross-domain generalisation.  Having multiple speakers produce each text
  (${\approx}8$ speakers/text) yields better out-of-domain transfer than
  maximising text coverage alone.  OOV rate is a poor proxy for WER given
  Whisper's pre-trained language modelling capacity.
\item \textbf{Utterance count:} 50--100 utterances per speaker suffice for
  near-saturation in-domain WER.
\end{itemize}

A remaining limitation is the real-to-synthetic gap: synthesised children's
voices, while intelligible, differ from actual children in prosody and
articulatory characteristics, and our results confirm that models trained solely
on synthetic data still under-perform real-data systems.  Future work should
explore TTS models explicitly tuned for children's voice characteristics, and
extend the analysis to clinical populations where data scarcity is most acute.

\bibliographystyle{IEEEtran}
\bibliography{references}

\end{document}